\documentclass[cits]{PoS}
\title{Lowering the background level and the energy threshold of Micromegas x-ray detectors for axion searches}
\ShortTitle{Micromegas x-ray detectors for axion searches}

\author{\speaker{F.J.~Iguaz},
        F.~Aznar \thanks{Present address: Centro Universitario de la Defensa,
        Universidad de Zaragoza, Zaragoza, Spain},
        J.F.~Castel, T.~Dafni, J.A.~Garcia, J.G.~Garza, I.G.~Irastorza,
        I.~Ortega\thanks{Present address: CERN, European Organization for Particle Physics and Nuclear Research.},
        A.~Rodr\'iguez and
        A.~Tomas\thanks{Present address: High Energy Physics group, Blackett Laboratory, Imperial College, London,
U.K.}\\
        Laboratorio de F\'isica Nuclear y Astropart\'iculas, Universidad de Zaragoza, Spain.\\
        E-mail: \email{iguaz@cern.ch}}
        
\author{S.~Aune, E.~Ferrer-Ribas, J.~Galan, I.~Giomataris, D.~Jourde and T.~Papaevangelou \\
        Service d`Electronique, des D\'etecteurs et d`Informatique, CEA-Saclay, 
        Gif-sur-Yvette, France}
        
\author{M.~Davenport and T.~Vafeiadis\\
        CERN, European Organization for Particle Physics and Nuclear Research,
        Geneva, Switzerland}
        
\author{S.C.~Yildiz\\
        Department of Physics, Dogus University, Acibadem Kadikoy, Istanbul, Turkey}

\abstract{Axion helioscopes search for solar axions by their conversion in x-rays in the presence of high magnetic fields.
The use of low background x-ray detectors is an essential component contributing to the sensitivity of these searches.
In this work, we review the recent advances on Micromegas detectors used in the CERN Axion Solar Telescope (CAST)
and proposed for the future International Axion Observatory (IAXO). The actual setup in CAST has achieved
background levels below 10$^{-6}$ keV$^{-1}$ cm$^{-2}$ s$^{-1}$, a factor 100 lower than the first generation
of Micromegas detectors. This reduction is based on active and passive shielding techniques, the selection
of radiopure materials, offline discrimination techniques and the high granularity of the readout.
We describe in detail the background model of the detector, based on its operation at CAST site
and at the Canfranc Underground Laboratory (LSC), as well as on Geant4 simulations. The best levels currently achieved
at LSC are low than 10$^{-7}$ keV$^{-1}$ cm$^{-2}$ s$^{-1}$ and show good prospects for the application
of this technology in IAXO. Finally, we present some ideas and results for reducing the energy threshold
of these detectors below 1 keV, using high-transparent windows, autotrigger electronics
and studying the cluster shape at different energies. As a high flux of axion-like-particles
is expected in this energy range, a sub-keV threshold detector could enlarge the physics case of axion helioscopes.}

\FullConference{Technology and Instrumentation in Particle Physics 2014\\
		 2-6 June, 2014\\
		 Amsterdam, the Netherlands}

\begin{document}
\section{Micromegas for axion searches}
Axion helioscopes {\bf\cite{Dafni:2014td}} aim to detect solar axions through their conversion by the inverse Primakoff effect
into x-rays (1-10 keV) using strong magnetic fields. The CERN Axion Solar Telescope (CAST) {\bf\cite{Zioutas:2005kz}}
is the most powerful example and has set the best exclusion limits in the axion-photon coupling
for a wide range of axion masses {\bf\cite{Andriamonje:2007sa,Arik:2014ea}}.
Three of the four CAST magnet bores are equipped with microbulk Micromegas (MM) detectors
{\bf\cite{Andriamonje:2010sa, Aune:2014sa}}
as they have a light and radiopure material budget {\bf\cite{Cebrian:2011sc}};
show an excellent energy resolution (12\% FWHM at 5.9 keV) {\bf\cite{Iguaz:2012fi}};
have a high discrimination power to select x-rays (point-like events) from muons and gammas;
can be shielded by applying the same techniques used in rare events experiments.

\medskip
X-ray detectors are an important issue for the future International Axion Observatory
(IAXO) {\bf\cite{Irastorza:2014,Armengaud:2014ea}}. Its main goal is to improve the signal-to-noise ratio more than
$10^4$ with respect to CAST, i.e. 1-2 orders of magnitude in sensitivity to the axion-photon coupling.
From the different improvements foreseen, low background x-rays detectors must reach values in
$10^{-7}$-$10^{-8}$ s$^{-1}$ keV$^{-1}$ cm$^{-2}$. MM technology is a perfect candidate according to
the latest level reached in CAST-MM during 2013 {\bf\cite{Garza:2014jg}} and currently the limit achieved
at the Canfranc Underground Laboratory (LSC).

\medskip
In section {\bf2}, we will briefly describe the actual CAST-MM background model (exposed in detail in {\bf\cite{Aune:2014sa})}.
We will thus present the physical motivation and the R\&D for sub-keV CAST-MM detectors in section {\bf 3} and,
we will finish with some conclusions and prospects.

\section{Background model of CAST-MM detectors}
In table {\bf\ref{tab:BackModel}} the different contributions to background model of CAST-MM detectors are detailed.
They have been identified either by in-situ measurements at CAST, by a detector's replica installed at LSC
or by Geant4 simulations. The main contribution is caused by muons, while the origin of actual limit set at LSC
is unknown. Some hypothesis are neutrons or the $^{39}$Ar isotope.

\begin{table}[htb!]
\begin{tabular}{c|cc|c}
{Contribution}&\multicolumn{2}{c|}{Level (s$^{-1}$ keV$^{-1}$ cm$^{-2}$)}&{Shielding technique applied}\\
{}&{Before}&{After}&{}\\
\hline
{Gamma flux}&{$7 \times 10^{-5}$}&{None?}&{Full coverage by 10 cm lead shielding}\\
{Radon}&{$8 \times 10^{-7}$}&{None}&{Nitrogen flux inside the shielding}\\
{Cosmic muons}&{$2 \times 10^{-6}$}&{$6 \times 10^{-7}$}&{95\% coverage by an active muon veto}\\
{Al cathode}&{$5 \times 10^{-7}$}&{None}&{Replacement by an ultrapure copper cathode}\\
\hline
{LSC limit}&\multicolumn{2}{c|}{$1.1 \times 10^{-7}$}&{Neutrons? $^{39}$Ar? Others?}
\end{tabular}
\caption{The different contributions to background level in 2-7 keV range identified in CAST-MM background model
and the different shielding techniques applied to completely or partially remove them. For clarity, values are given
without errors. The full measurements with the associated errors are in {\bf\cite{Aune:2014sa}}.}
\label{tab:BackModel}
\end{table}

\section{Lowering the energy threshold}
\begin{figure}[htb!]
\centering
\includegraphics[width=65mm]{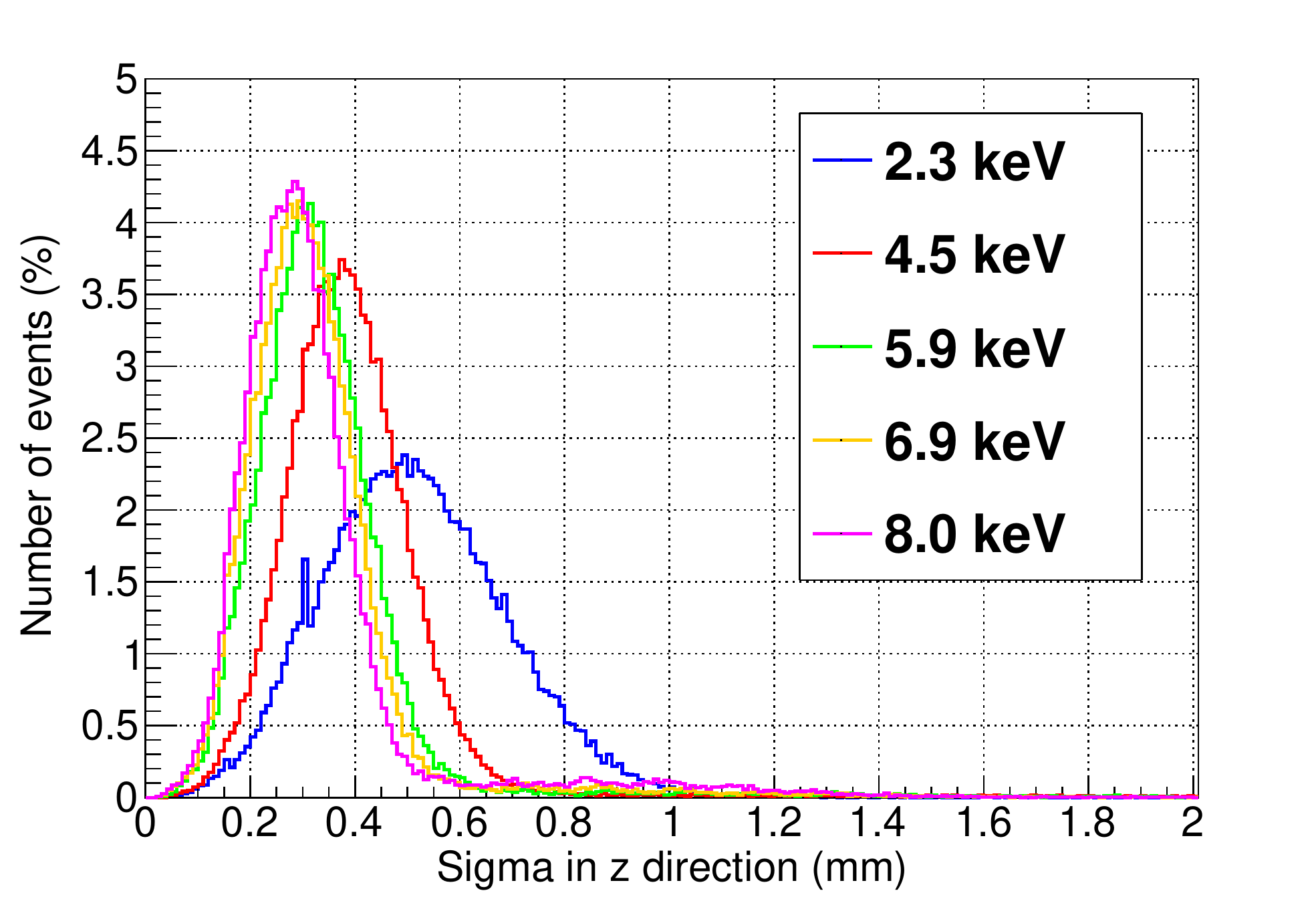}
\includegraphics[width=65mm]{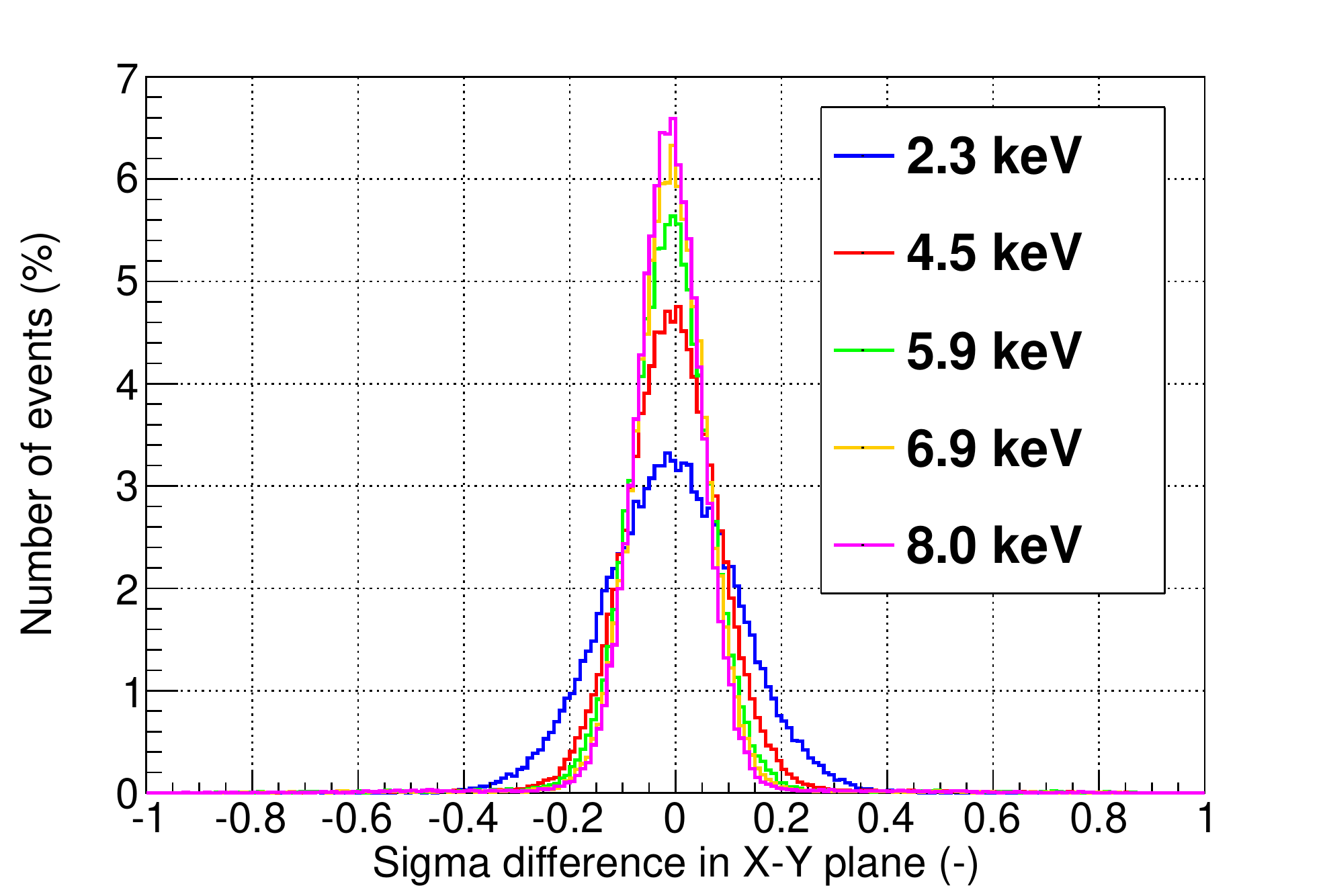}
\caption{Distribution of the cluster's width in Z (left, in sigma units) and
the cluster's width difference in XY (right) for the different fluorescence lines used
in the x-ray calibration of CAST-M18.}
\label{fig:Xrays}
\end{figure}

New calculations of axion production at the Sun by the so-called ``BCA processes'' in non-hadronic axion
models {\bf\cite{Redondo:2013jr}} point out to an axion flux peaking at energies around 1 keV. This fact motivates
the use of sub-keV detectors in IAXO. With this aim, there is a R\&D line based on:

\begin{itemize}
 \item[{\bf Window}] The x-rays coming from the magnet enter the conversion volume via a gas-tight window made of 5 $\mu$m
 aluminized mylar foil. This foil is supported by a metallic squared-pattern strong-back, to withstand
 the pressure difference to the magnet's vacuum system. This foil is transparent down to energies of 1.5 keV.
 Other possible materials like polyimide-based foils are being studied, as their transparency is enlarged down to 0.7 keV.
 \item[{\bf Gas}] Other mixtures than Ar+2\%isobutane have been studied to increase
 the actual operation point ($8 \times 10^3$), i.e, to decrease the energy threshold. A value of $3 \times 10^4$ is reached
 in neon-based mixtures but the pressure must be increased to maximize the photon conversion {\bf\cite{Iguaz:2012fi}}.
 \item[{\bf Electronics}] The actual one is based on the AFTER chip {\bf\cite{Baron:2008pb}}, which provides time information
 of each strip. This feature improves the signal-to-noise ratio and have reduced the energy threshold down to 450 eV.
 Autotrigger systems like AGET {\bf\cite{Anvar:2011sa}} may further enhance this reduction.
 \item[{\bf Analysis}] The CAST-MM detectors have been calibrated in an electron beam at the CAST Detector Laboratory
 {\bf\cite{Vafeiadis:2013tv}}.
 In this setup, the fluorescence lines of different target materials ranging from 2.3 (gold) to 8.0 keV (copper)
 are used to calculate the integrated signal efficiency {\bf\cite{Garza:2014jg}}. This data also provides information
 on event's topology, which can be used in the analysis. To illustrate this issue, the distribution
 of the cluster's width in Z (left) and the cluster difference between X\&Y widths (right)
 are shown respectively in figure {\bf\ref{fig:Xrays}}.
 Clusters are wider at low energies because most of the x-rays may be absorbed in the first milimeters
 just after the window and will suffer more diffusion effects. Cluster differences also
 increase at low energies as charge fluctuations between the two detector planes (XY) are more important.
\end{itemize}

\section{Conclusions and prospects}
During more than a decade in CAST, Micromegas detectors have reduced their background levels by two orders of magnitude,
reaching a value of $~7 \times 10^{-7}$ keV$^{-1}$ cm$^{-2}$ s$^{-1}$. Most of contributions
have been identified and removed applying shielding techniques used in rare event experiments. The best levels currently
achieved at LSC (a factor 7 lower) show good prospects for the application of this technology in IAXO. Recent calculations
of axion flux motivate the use of sub-keV detectors in CAST. We have presented the main open R\&D lines on this issue
for CAST-MM.

%\section*{Acknowledgements}
\acknowledgments
We thank to our colleagues of CAST for many years of collaborative work, R. de Oliveira and his team at CERN
for the manufacturing of the microbulk readouts and the LSC staff for their support during the use of their facilities.
We acknowledge the support from the European Commission under the European Research Council T-REX Starting Grant
ref. ERC-2009-StG-240054 of the IDEAS program of the 7th EU Framework Program.
We also the acknowledge support from the Spanish Ministry MINECO under contracts ref. FPA2008-03456 and FPA2011-24058,
as well as under the CPAN project ref. CSD2007-00042 from the Consolider-Ingenio 2010 program.
These grants are partially funded by the European Regional Development funded (ERDF/FEDER).
My personal gratitude is for the support from the \emph{Juan de la Cierva} program.

\end{document}